\documentclass[preprint,superscriptaddress,nofootinbib]{revtex4}
\usepackage[dvips]{graphicx}
\usepackage[usenames]{color}
\usepackage{setspace}
\usepackage{amsmath}
\usepackage{amssymb}
\usepackage{amsthm}
\usepackage{epsf}
\usepackage{epsfig}

\newcommand{\GeV}{{\text{GeV}}}
\newcommand{\eff}{{\text{eff}}}

\begin{document}


\preprint{UT-HET-117}
\preprint{KIAS-P16065}

\title{
  Gravitational waves and Higgs boson couplings for exploring first
  order phase transition in the model with a singlet scalar field
}
\author{Katsuya~Hashino}
\email{hashino@jodo.sci.u-toyama.ac.jp}
\affiliation{
  Department of Physics,
  University of Toyama, 3190 Gofuku, Toyama 930-8555, Japan
}
\author{Mitsuru~Kakizaki}
\email{kakizaki@sci.u-toyama.ac.jp}
\affiliation{
  Department of Physics,
  University of Toyama, 3190 Gofuku, Toyama 930-8555, Japan
}
\author{Shinya~Kanemura}
\email{kanemu@sci.u-toyama.ac.jp}
\affiliation{
  Department of Physics,
  University of Toyama, 3190 Gofuku, Toyama 930-8555, Japan
}
\author{Pyungwon~Ko}
\email{pko@kias.re.kr}
\affiliation{
  School of Physics, KIAS, Seoul 02455, Korea
}
\author{Toshinori~Matsui}
\email{matsui@kias.re.kr}
\affiliation{
  School of Physics, KIAS, Seoul 02455, Korea
}


\begin{abstract}

We calculate the spectrum of gravitational waves originated 
from strongly first order electroweak phase transition 
in the extended Higgs model with a real singlet field. 
In order to calculate the bubble nucleation rate, we perform a two-field 
analysis to evaluate bounce solutions connecting the true and the false vacua 
using the one-loop effective potential at finite temperatures. 
Imposing the Sakharov condition of the departure from thermal equilibrium 
for baryogenesis, we survey allowed regions of parameters of 
the model. 
We then investigate the gravitational waves produced at electroweak 
bubble collisions in the early Universe, 
such as the sound wave, the bubble wall collision and the plasma turbulence. 
We find that the strength at the peak frequency can be large enough
to be detected at future space-based 
gravitational interferometers such as eLISA, DECIGO and BBO.
Predicted deviations in the various Higgs boson couplings are also evaluated 
at the zero temperature,
and are shown to be large enough too.
Therefore, in this model strongly first order electroweak phase transition 
can be tested by the combination 
of the precision study of various Higgs boson couplings at LHC, 
the measurement of the triple Higgs boson coupling at future 
lepton colliders and the shape of the spectrum of gravitational wave 
detectable at future gravitational interferometers. 

\end{abstract}


\maketitle



By the discovery of a Higgs boson~\cite{Aad:2012tfa,Chatrchyan:2012xdj} and 
the dedicated measurements of its property at LHC, 
the mass generation mechanism for elementary particles 
in the standard model (SM) has been established. 
One of the next important targets of high energy physics is 
to explore the structure of the Higgs sector, 
dynamics of electroweak symmetry breaking (EWSB), 
and the nature of the Higgs boson ($h$).


In addition, the mechanism of electroweak phase transition (EWPT)
is still a mystery, 
that is strongly related not only to the physics behind EWSB 
but also to various cosmological problems 
such as baryon asymmetry of the Universe and 
cosmic inflation. 
In particular, the strongly first order phase transition (1stOPT) 
is crucial for a successful scenario of electroweak baryogenesis 
(EWBG)~\cite{Kuzmin:1985mm}. 
With another requirement of additional CP violating phase, 
the EWBG scenario can be realized by introducing an extended Higgs sector.  
Therefore, in this scenario, the physics of EWBG
can be in principle tested by exploring the Higgs sector.


It has been well known that the 1stOPT is realized 
by the non-decoupling thermal loop effects 
on the finite temperature effective potential and/or by the field mixing 
of the Higgs boson with additional scalar 
fields~\cite{Funakubo:1993jg,Cline:1996mga,Kanemura:2004ch,Funakubo:2005pu,Espinosa:2007qk,Profumo:2007wc,Noble:2007kk,AKS,Kanemura:2011fy,Gil:2012ya,Fairbairn:2013uta,Li:2014wia,Tamarit:2014dua,Kanemura:2014cka,Fuyuto:2014yia,Profumo:2014opa,Chao:2014ina,Blinov:2015vma,Fuyuto:2015vna,Karam:2015jta,Kakizaki:2015wua,Hashino:2016rvx}.  
These effects also affect the effective potential at the zero temperature, 
so that they normally deviate the triple Higgs boson coupling 
(the $hhh$ coupling) typically by larger than 
$10\%$~\cite{Kanemura:2004ch,Noble:2007kk,AKS,Kanemura:2011fy,Hashino:2015nxa,Tamarit:2014dua,Kanemura:2014cka,Kakizaki:2015wua,Hashino:2016rvx}. 
It may be challenging for the (high luminosity) LHC 
to achieve this level of accuracy.
However, the plan of the International Linear Collider (ILC)~\cite{ILC}
includes the determination of the $hhh$ coupling
with $10\%$ accuracy by upgrading the center-of-mass energy
to $\sqrt{s}=1~{\rm TeV}$~\cite{ILCHiggsWhitePaper,Moortgat-Picka:2015yla,Fujii:2015jha}.
The Compact LInear Collider (CLIC)~\cite{CLIC} 
also aims to reach the similar accuracy.
The Future Circular Collider of electrons and positrons (FCC-ee)~\cite{FCC-ee}
will not address the precision measurement of the $hhh$ coupling
as its center-of-mass energy is insufficient.
The possibility of testing the $hhh$ coupling 
at future hadron colliders with $\sqrt{s} = 100~{\rm TeV}$ is
also considered~\cite{He:2015spf}. 
Therefore, the scenario of EWBG
can be tested by precision measurements of the $hhh$ coupling 
at future collider experiments.  
In a class of models where the 1stOPT is caused by 
the field mixing, resulting predicted values for the Higgs boson couplings 
such as those with weak gauge bosons and with fermions can also be 
deviated significantly because of the field mixing. 
Therefore, this class of models for EWBG is expected 
to be tested by the data from LHC not only those at future linear colliders.


On the other hand, it has also been known that 
strongly 1stOPT at the early Universe 
is a discriminative origin of gravitational waves 
(GWs)~\cite{Apreda:2001us,Grojean:2006bp,Espinosa:2008kw,Kakizaki:2015wua,Caprini:2015zlo,Hashino:2016rvx,Kobakhidze:2016mch}.  
Recently, the GWs has been directly detected 
at the Advanced LIGO experiment which has an astronomical 
origin~\cite{aLIGO}. 
By this discovery, measurements of GWs with various 
frequencies will be accelerated in the near 
future including KAGRA~\cite{Somiya:2011np}, 
Advanced LIGO~\cite{Harry:2010zz} and Advanced VIRGO~\cite{Accadia:2009zz}, 
by which new field of GW astronomy will be 
extensively developed.  
Furthermore, future space based GW interferometers such as 
eLISA~\cite{Seoane:2013qna}, DECIGO~\cite{Kawamura:2011zz}
and BBO~\cite{Corbin:2005ny} provide us an opportunity of measuring GWs 
with a wider range of frequencies, which can cover GWs from 
the first order EWPT.
Therefore, by precisely measuring the spectrum of GWs, we 
can test the physics of EWPT and further 
the scenario of EWBG.


In this Letter, we calculate the spectrum of GWs originated 
from strongly first order EWPT in a concrete renormalizable model, 
the extended Higgs model with a real singlet field. 
In order to calculate the bubble nucleation rate, 
we perform a two-field analysis to evaluate bounce solutions 
connecting the true and the false vacua 
using the one-loop effective potential at finite temperatures. 
We survey allowed regions of parameters of the model imposing the 
Sakharov condition of the departure 
from thermal equilibrium for baryogenesis~\cite{Sakharov:1967dj}. 
We then investigate the GWs produced at 
electroweak bubble collisions in the early Universe, 
such as the sound wave, the bubble wall collision and the plasma turbulence. 
We find that in this model strongly first order EWPT
can be well tested by the combination 
of the precision study of various Higgs boson couplings at LHC, 
the measurement of the $hhh$ coupling at future 
lepton colliders and the spectrum of GWs detectable at eLISA and DECIGO.


Let us begin with a brief review of
the Higgs singlet model (HSM), which is one of the simplest extensions of the 
SM~\cite{Profumo:2007wc,Noble:2007kk,Ashoorioon:2009nf,Espinosa:2011ax,Fuyuto:2014yia,Chen:2014ask,Kanemura:2015fra,Kanemura:2016lkz}.  
The Higgs sector of the HSM is equipped with a real isospin scalar singlet $S$
in addition to the Higgs doublet $\Phi$.
The general tree-level Higgs potential allowed by gauge invariance
and renormalizability is given by
\begin{align}
  V_0=-\mu^2_\Phi |\Phi|^2+\lambda_\Phi |\Phi|^4
  +\mu_{\Phi S} |\Phi|^2S+\frac{\lambda_{\Phi S}}{2} |\Phi|^2S^2
  +\mu^3_SS+\frac{m^2_S}{2}S^2+\frac{\mu^{\prime}_S}{3}S^3+\frac{\lambda_S}{4}S^4,
\label{treepotential}
\end{align} 
with eight parameters
$\mu^2_\Phi, m^2_S, \lambda_\Phi^{}, \lambda_S^{}, \lambda_{\Phi S}^{},
\mu_{\Phi S}^{}, \mu^{\prime}_S$ and $\mu^3_S$.
\footnote{One of the mass parameters can be removed by the field redefinition
  of the singlet field without loss of 
generality~\cite{Espinosa:2011ax,Chen:2014ask}.}
After the condensation of the two Higgs fields, they are expanded
around the vacuum expectation values $v_\Phi^{}$ and $v_S^{}$ as
\begin{align}
  \Phi= \left(
  \begin{array}{c}
  G^+ \\
  \frac{1}{\sqrt{2}}(v_\Phi^{} + \phi_1^{} + i G^0)
  \end{array}
  \right) , \quad S= v_S^{} + \phi_2^{}.
\end{align}
There appear two physical degrees of freedom $\phi_1^{}$ and
$\phi_2^{}$ that mix with each other in addition to
Nambu-Goldstone (NG) modes $G^\pm$ and $G^0$ that are absorbed 
by the $W$- and $Z$-bosons.
In the following, we analyze the phase structure of this HSM 
in the classical field space spanned by
\begin{align}
  \langle \Phi \rangle = \left(
  \begin{array}{c}
  0 \\
  \frac{1}{\sqrt{2}}\varphi_\Phi^{}
  \end{array}
  \right) , \quad \langle S \rangle = \varphi_S^{}.
\end{align}


Radiative corrections modify the shape of the Higgs potential from
the tree-level form.
At zero temperature, the effective potential up to the one-loop level 
is~\cite{Coleman:1973jx}
\begin{align}
  V_{\eff, T=0}^{}(\varphi_\Phi^{},\varphi_S^{})
  =V_0(\varphi_\Phi^{},\varphi_S^{}) +\sum_i n_i^{} \ 
  \frac{M^4_i(\varphi_\Phi^{},\varphi_S^{})}{64\pi^2}
  \left(\ln\frac{M^2_i(\varphi_\Phi^{},\varphi_S^{})}{Q^2} -c_i \right),
\end{align}
where $Q$ is the renormalization scale, which is set at $v_\Phi^{}$
in our analysis.
Here, $n_i$ and $M_i(\varphi_\Phi,\varphi_S)$ denote the degrees of the freedom
and the field-dependent masses for particles $i$, respectively.
We take the $\overline{\rm MS}$ scheme, where the numerical constants $c_i$ are
set at $3/2$ ($5/6$) for scalars and fermions (gauge bosons).
We impose the tadpole conditions using the one-loop level 
effective potential as
$\left\langle \partial V_{\eff, T=0}^{}/\partial \varphi_i^{} \right\rangle=0$,
with $i=\Phi$ or $S$.  
Here, the angle bracket $\langle\cdots\rangle$ represents
the field-dependent quantity evaluated at
our true vacuum $(\varphi_\Phi^{}, \varphi_S^{}) = (v_\Phi^{}, v_S^{})$.
The mass squared matrix of the real scalar bosons 
in the $(\phi_1, \phi_2)$ basis is diagonalized as
\begin{align}
  m_{i j}^2 = \left\langle 
  \frac{\partial^2 V_{\eff, T=0}^{}}{\partial \varphi_i^{} \partial \varphi_j^{}}
  \right\rangle
=
\begin{pmatrix}
\cos \theta & -\sin \theta \\
\sin \theta & \cos \theta
\end{pmatrix}
\begin{pmatrix}
m_h^2 & 0 \\
0 & m_H^2
\end{pmatrix}
\begin{pmatrix}
\cos \theta & \sin \theta \\
-\sin \theta & \cos \theta
\end{pmatrix},
\end{align}
leading to one-loop improved mass eigenvalues of the Higgs bosons
$m_h$ and $m_H$, and mixing angle $\theta$ ($m_h^{} < m_H^{}$,
$- \pi/4 \leq \theta \leq \pi/4$). 
The lighter boson $h$ is identified with the discovered Higgs boson with the mass 
$125\GeV$ in this Letter, and the alternative case where $H$ is the discovered one
will be examined elsewhere.
From the above equations, we use
$v_\Phi^{}$, $v_S^{}$, $m_h^{}$, $m_H^{}$ and $\theta$ as the input parameters 
instead of $\mu_\Phi^2$, $m_S^2$, $\lambda_\Phi^{}$, $\mu_{\Phi S}^{}$ 
and $\mu'_S$.


Due to finite temperature effects, the effective potential is modified to~\cite{Dolan:1973qd}
\begin{align}
  V_{\eff,T}^{}[M_i^2(\varphi_\Phi^{},\varphi_S^{})]
  =V_{\eff, T=0}^{}(\varphi_\Phi^{},\varphi_S^{})
  + \sum_i n_i \ \frac{T^4}{2\pi^2}I_{B,F} 
  \left( \frac{ M^2_i(\varphi_\Phi^{},\varphi_S^{})}{T^2} \right),
\end{align}
where 
\begin{align}
  I_{B,F}(a^2)= \int^{\infty}_0 dx \ x^2 \ln 
  \left(1 \mp \exp^{-\sqrt{x^2+a^2}} \right) ,
\end{align}
for boson and fermions, respectively.
In order to take ring-diagram contributions into account,
we replace the field-dependent masses in the effective potential
as~\cite{Carrington:1991hz}
\begin{align}
  M_i^2(\varphi_\Phi^{},\varphi_S^{})
  \to M_i^2(\varphi_\Phi^{},\varphi_S^{}, T)
  = M_i^2(\varphi_\Phi^{},\varphi_S^{})+\Pi_i(T),
\end{align}
where $\Pi_i^{}(T)$ stand for the finite temperature contributions to
the self energies.
We consider loop contributions from the fields
$i=h, G^\pm, G^0, H, W_{T, L}^{\pm}, Z_{T, L}^{}, \gamma_{T, L}^{}, t$ and $b$.
As for the scalar sector particles, the thermally corrected field-dependent
masses are given by~\cite{Ashoorioon:2009nf}
\begin{align}
  M^2_{h, H}(\varphi_\Phi^{},\varphi_S^{},T)=
  &\frac{1}{2}\left(M_{11}^2+M_{22}^2 \mp 
    \sqrt{(M_{11}^2-M_{22}^2)^2 + 4 M_{12}^2 M_{21}^2} \right), \\
  M^2_{G^0, G^\pm}(\varphi_\Phi^{},\varphi_S^{},T)=
  &-\mu_\Phi^2 + \lambda_\Phi^{} \varphi_\Phi^2 + \mu_{\Phi S}^{} \varphi_S
    + \frac{\lambda_{\Phi S}^{}}{2}\varphi_S^2 \nonumber\\
  &+\frac{T^2}{48}(9 g^2+3 g'^2+12 (y_t^2+y_b^2) 
    + 24 \lambda_\Phi^{} + 2 \lambda_{\Phi S}^{}), 
\end{align}
where
\begin{align}
  \begin{pmatrix}
    M_{11}^2 & M_{12}^2 \\
    M_{21}^2 & M_{22}^2
  \end{pmatrix}
               = &
  \begin{pmatrix}
    -\mu_\Phi^2 + 3\lambda_\Phi^{} \varphi_\Phi^2
    + \mu_{\Phi S}^{} \varphi_S^{} + \frac{\lambda_{\Phi S}^{}}{2} \varphi_S^2
    & \mu_{\Phi S}^{} \varphi_\Phi^{} 
    + \lambda_{\Phi S}^{} \varphi_\Phi^{} \varphi_S^{} \\
    \mu_{\Phi S}^{} \varphi_\Phi^{} 
    + \lambda_{\Phi S}^{} \varphi_\Phi^{} \varphi_S^{} 
    & m_S^2 + 2\mu'_S \varphi_S^{} + 3\lambda_S^{} \varphi_S^2
    +\frac{\lambda_{\Phi S}^{}}{2}\varphi_\Phi^2
  \end{pmatrix}\nonumber\\
&+\frac{T^2}{48}
	\begin{pmatrix}
	9 g^2+3 g'^2+12 (y_t^2+y_b^2) + 24 \lambda_\Phi^{} 
        + 2 \lambda_{\Phi S}^{} & 0 \\
	0 & 12 \lambda_S^{} + 8 \lambda_{\Phi S}^{}
	\end{pmatrix}. 
\end{align}
Here, $g$ and $g'$ ($y_t^{}$ and $y_b^{}$) represent 
the $SU(2)$, $U(1)$ gauge coupling constants (the top and bottom Yukawa coupling constants).
For the thermal corrections to the field-dependent masses of the EW 
gauge bosons, see, for example, Ref.~\cite{Hashino:2016rvx}.
On the other hand, fermion counterparts do not receive such thermal 
corrections.


Before analyzing the phase structure utilizing the finite temperature 
effective potential, let us briefly summarize
important theoretical and experimental constraints on the HSM.
In order to retain perturbative unitarity, 
the absolute values of eigenvalues of $S$-wave 
scattering amplitudes for the weak gauge bosons and scalars
should be smaller than $1/2$~\cite{Lee:1977eg}.
This constraint is converted to the inequality
$m_{h}^2 \cos^2 \theta + m_{H}^2 \sin^2 \theta \leq 4 \pi \sqrt{2}/(3 G_F)
\simeq \left( 700~\GeV \right)^2$.
Since $m_h^{}$ has been measured, 
an upper bound on the mixing angle $\theta$ 
is obtained as a function of $m_H^{}$.
In order for the Higgs potential to be bounded from below, 
the vacuum stability condition has to be satisfied at a scale 
$\mu$~\cite{Fuyuto:2014yia}:
\begin{align}
\lambda_\Phi^{}(\mu)>0, \quad
\lambda_{S}^{}(\mu)>0, \quad
4\lambda_{\Phi}^{}(\mu)\lambda_S^{}(\mu)>\lambda^2_{\Phi S}(\mu).
\label{eq:stability}
\end{align}
It should be also noticed that 
in general the Higgs potential has several local minima.
In order to prevent our EW phase from decaying into another one,
the EW phase needs to be the global minimum of the Higgs potential 
as~\cite{Chen:2014ask}
\begin{align}
  V_{\eff, T=0}^{}({\rm EW~phase})<V_{\eff, T=0}^{}({\rm other~phases}). 
\label{eq:globalmin}
\end{align}
Although the couplings in the Higgs potential 
remain perturbative at the EW scale, 
they become strong at higher energy scales due to renormalization group
flow.
In this Letter,
the Landau pole $\Lambda^{}$ is defined
as the scale where any of the Higgs couplings is as strong as
$|\lambda_{\Phi,S,\Phi S}(\Lambda)|=4\pi$~\cite{Kanemura:2016lkz}.
For the HSM with strongly 1stOPT, the Landau pole ranges typically
from a few TeV to around $10$~{\rm TeV} depending on parameter 
choices~\cite{Fuyuto:2014yia}.
The introduction of the scalar singlet $S$ also affects
the self-energies of the $W$- and $Z$-bosons.
For the details of the computations of the oblique parameters,
see Ref.~\cite{Baek:2011aa}.
Given the observed value of the Higgs boson mass 
$m_h^{} \simeq 125~\GeV$, 
the Higgs boson mixing angle is constrained as
$\cos \theta \gtrsim 0.92$ for $m_H^{} \gtrsim 400~\GeV$~\cite{Baek:2012uj}.


The presence of the scalar singlet gives rise to
deviations in the couplings of the discovered 
Higgs boson from their SM values.
The dominant contributions to the deviations are induced by
the mixing between the two Higgs bosons.
Therefore, the predicted Higgs boson couplings 
to the gauge boson $V=W^\pm, Z$ and fermions $F$ normalized 
by the corresponding SM ones are universal as
$\kappa = \kappa_V^{} = \kappa_F^{} =\cos \theta$.
The most stringent bounds are extracted
from the measurements of the Higgs boson decay into weak gauge bosons
at the LHC Run-I as
$\kappa_Z^{}=1.03^{+0.11}_{-0.11}, \kappa_W^{}=0.91^{+0.10}_{-0.10}$~\cite{kappa}.
The high-luminosity stage of the LHC can reach the precision
of $2\%$~\cite{CMS:2013xfa}.
The precision of the Higgs boson coupling measurements will be
significantly improved once electron-positron colliders are realized.
In the case of the ILC with $\sqrt{s}=500~{\rm GeV}$, 
the expected accuracy can be $0.37\%$ ($0.51\%$) for $\kappa_Z^{}$ 
($\kappa_W^{}$)~\cite{Fujii:2015jha}.


The value of the $hhh$ coupling $\lambda_{hhh}^{}$
is considerably altered
by the new singlet scalar.
Here, for simplicity, we adopt the effective potential approach
in computing $\lambda_{hhh}^{}$.
\footnote{Dependence of the $hhh$ coupling
on the external momentum is discussed in 
Refs.~\cite{Kanemura:2002vm,Kanemura:2004mg}.}
In this approximation, the SM prediction is obtained as
\begin{align}
  \lambda_{hhh}^{\rm SM}  &= \frac{3m_{h}^2}{v_\Phi}
  \left[
	1+\frac{9m_{h}^2}{32\pi^2v_\Phi^2}+\sum_{i=W^\pm,Z,t,b}n_i\frac{m_i^4}{12\pi^2m_{h}^2v_\Phi^2}
\right]\simeq 176 \GeV , 
\end{align}
while the HSM counterpart is symbolically expressed as
\footnote{In our analysis at zero temperature, we disregard
the minor loop contributions from the NG modes
in order to avoid complexity.}
\begin{align}
\lambda_{hhh}^{\rm HSM}&=
c_\theta^3\left\langle\frac{\partial^3 V_{\eff, T=0}}{\partial \varphi_\Phi^3}\right\rangle
+c_\theta^2s_\theta\left\langle\frac{\partial^3 V_{\eff, T=0}}{\partial \varphi_\Phi^2\partial \varphi_S}\right\rangle
+c_\theta s_\theta^2\left\langle\frac{\partial^3 V_{\eff, T=0}}{\partial \varphi_\Phi\partial \varphi_S^2}\right\rangle
+s_\theta^3\left\langle\frac{\partial^3 V_{\eff, T=0}}{\partial \varphi_S^3}\right\rangle, 
\end{align}
where $c_\theta=\cos\theta$ and $s_\theta=\sin\theta$. 
In discussing the deviation in the $hhh$ coupling,
we exclusively utilize the following normalized quantity:
\begin{align}
\Delta \lambda_{hhh}^{}
= \frac{\lambda_{hhh}^{\rm HSM}-\lambda_{hhh}^{\rm SM} }{\lambda_{hhh}^{\rm SM}}.
\label{Eq.hhh}
\end{align}
So far, the LHC has set no meaningful constraint on the $hhh$
coupling.
In the future, at the high luminosity LHC with
$L=3000~{\rm fb}^{-1}$ 
the production cross section of the double Higgs production process
can be measured with $54\%$~\cite{LHChhh}.
Once realized, the ILC is capable of 
measuring the $hhh$ coupling with considerable accuracy.
The ILC stage with $\sqrt{s}=500~\GeV$ and $L=4000~{\rm fb}^{-1}$, 
the expected precision is $27\%$~\cite{Fujii:2015jha}.
At the ILC with the higher energy of $\sqrt{s}=1~{\rm TeV}$, 
the precision will be ameliorated to $16\%$ ($10\%$)
for $L=2000~{\rm fb}^{-1}$ ($L=5000~{\rm fb}^{-1}$)~\cite{Fujii:2015jha}. 

Let us consider the EWPT in the HSM.
In order for baryogenesis to work, the departure from 
thermal equilibrium must be realized.
In EWBG scenarios, 
the baryon number changing sphaleron process must decouple
quickly after the EWSB:
the sphaleron interaction rate $\Gamma_{\rm sph}^{}(T)$ is smaller than
the Hubble expansion rate $H(T)$.
This criterion is satisfied if the EWSB is of strongly 
first order
\begin{align}
\frac{\varphi_c^{}}{T_c^{}} > \zeta_{\rm sph}^{}(T_c^{}),
\label{eq:sph}
\end{align}
where $\varphi_c$ is the VEV for the true vacuum at
the critical temperature $T_c$.
The value of $\zeta_{\rm sph}^{}(T_c)$ is typically around unity, 
and it is estimated as $\zeta_{\rm sph}(T_c)=1.1-1.2$
for the HSM~\cite{Fuyuto:2014yia}.


Let us turn our discussion to GWs originated from the first order
EWPT in the HSM.
First, we introduce two important quantities $\alpha$ and $\beta$ 
that describe the dynamics of vacuum bubbles~\cite{Grojean:2006bp}.
First, we define the transition temperature such that
the bubble nucleation probability per Hubble volume per Hubble time
reaches the unity, $\Gamma/H^4|_{T=T_t^{}}^{}=1$. 
The parameter $\alpha$ is the ratio of the released energy density $\epsilon$
to the radiation energy density 
$\rho_{\rm rad}^{} =(\pi^2/30) g_*^{} T^4$ at the transition temperature $T_t^{}$:
\begin{align}
  \alpha \equiv \frac{\epsilon(T_t)}{\rho_{\rm rad}(T_t)}.
\end{align}
In our analysis, for simplicity, we neglect the temperature dependence of
the relativistic degrees of freedom, and fix it at $g_*^{}=107.65$. 
The parameter $\beta$ is the inverse of the time variation scale
of the bubble nucleation rate $\Gamma(t)=\Gamma_0^{}\exp(\beta t)$.
It is a standard to use the normalized
dimensionless parameter $\widetilde{\beta}$, which is defined as
\begin{align}
  \widetilde{\beta} \equiv \frac{\beta}{H_t^{}}
 = T_t^{} \frac{d}{d T}\left(\frac{S_3^{}(T)}{T}\right)\Bigg|_{T=T_t}^{},
\end{align}
where $S_3^{}(T)$ is the three dimensional Euclidean action 
for the bounce configuration of the classical field that
connects the true and the false vacua at $T$.
Once $T_t^{}$, $\alpha$ and $\beta$ are computed,
one can estimate the spectrum of the stochastic GWs
using the approximate analytic formula provided in Ref.~\cite{Caprini:2015zlo}.


We are now at the stage of discussing the testability
of the HSM by utilizing the interplay of measurements
of Higgs boson couplings at future colliders
and of GWs at future space-based interferometers.
In the light of the constraints on the HSM discussed above,
we perform numerical analysis in computing the Higgs boson couplings
and bubble dynamics parameters. 
The characteristic feature of the HSM (without $Z_2^{}$ symmetry) is that
the potential barrier between the true and the false vacua necessary 
for 1stOPT can be formed mainly by tree-level interactions, 
in sharp contrast to
the two Higgs doublet models or $Z_2^{}$ symmetric HSM, where non-decoupling loop effects
are necessitated for 1stOPT.
As an example scenario where
strongly 1stOPT is accomplished due to
large doublet-singlet Higgs mixing parameters 
$\mu_{\Phi S}^{}$ and $\lambda_{\Phi S}^{}$,
we consider the benchmark point shown in Table~\ref{tab:benchmark}
\footnote{ For the purpose of 
straightforward comparison, we take the same benchmark point 
as Case (ii) in Ref.~\cite{Fuyuto:2014yia}. }. 
In evaluating $T_t, \alpha$ and $\tilde{\beta}$, 
we implement the HSM into the public code 
{\tt CosmoTransitions}~\cite{Wainwright:2011kj}, which computes quantities
related to the cosmological phase transition in the multi-field space.
Predicted GW spectra are estimated using the approximate
analytic formula developed in \cite{Caprini:2015zlo,Espinosa:2010hh}.

\begin{table}[t]
  \center
  \caption{The benchmark point and scanned range for the HSM parameters.}
  \label{tab:benchmark}
  \begin{tabular}{|c|c|c|c|c|c||c|c|}
    \hline
    $v_\Phi^{}$~[GeV] & $v_S^{}$~[GeV] & $m_h^{}$~[GeV] & $\mu_{\Phi S}^{}$~[GeV] & $\mu_S^\prime$~[GeV] & $\mu_S^{}$~[GeV] & $m_H^{}$~[GeV] & $\theta$~[degrees] \\
    \hline\hline
    $246.2$ & $90$ & $125.5$ & $-80$ & $-30$ & $0$ & [$160$, $240$] & [$-45$, $0$] \\
\hline
  \end{tabular}
\end{table}

\begin{figure}[t]
  \centerline{\includegraphics[width=0.8\textwidth]{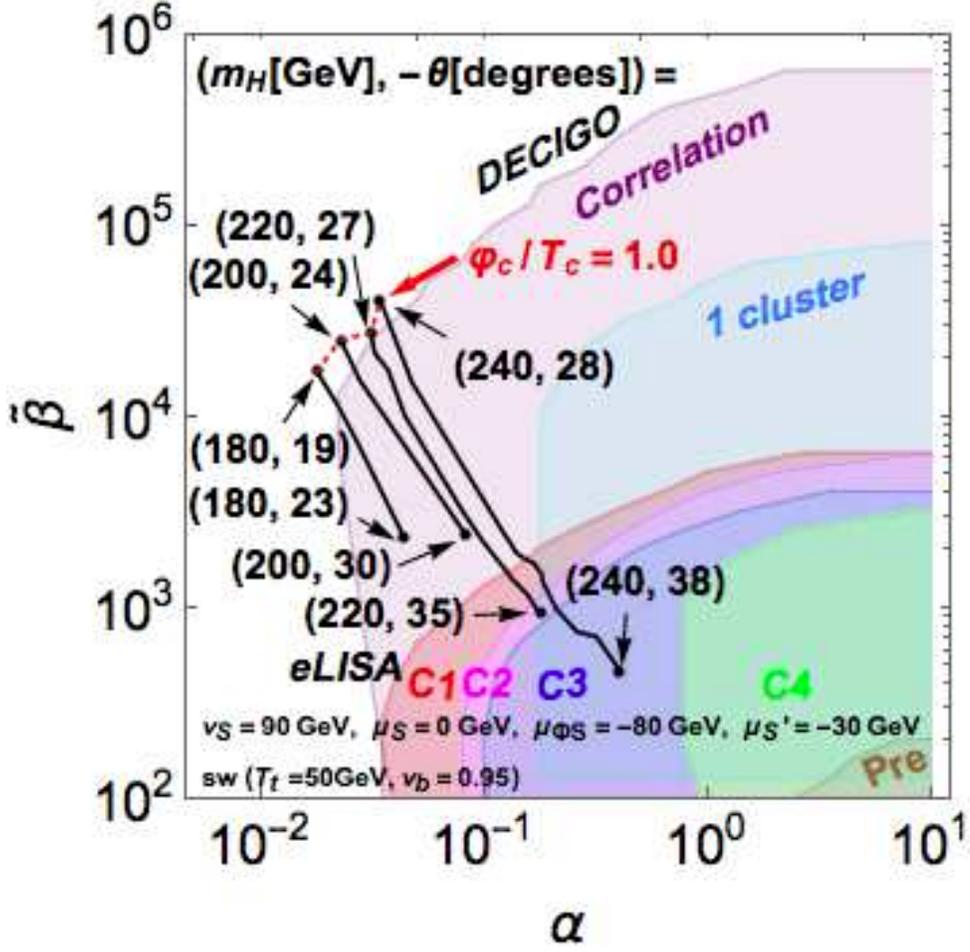}}
  \caption{The predicted values of $\alpha$ and $\widetilde{\beta}$
    with the variation of ($m_H^{}$, $- \theta$)
    in the HSM for the benchmark point in Table~\ref{tab:benchmark}.
    The expected sensitivities of eLISA and DECIGO detector configurations
    are set by using the sound wave contribution 
    for $T_t^{}=50~\GeV$ and $v_b^{}=0.95$. }
  \label{fig:GW1}
\end{figure}

In Fig.~\ref{fig:GW1}, we present the predicted values of $\alpha$
and $\beta$ with the variation of ($m_H^{}$, $- \theta$).
in the HSM for the benchmark point in Table~\ref{tab:benchmark}. 
The black curves show the predicted values of $\alpha$ and $\beta$
for $m_H^{}=180~\GeV$, $200~\GeV$, $220~\GeV$ and $240~\GeV$ 
from the left.
The upper bound on $\tilde{\beta}$ is set by the condition 
$\varphi_c^{}/T_c^{}=1$.
Since $S_3^{}(T)/T$ is a concave function of $T$,
not only positive but also negative values of $\widetilde{\beta}$ may be derived
from the criterion 
$\Gamma/H^4|_{T=T_t{}}^{}=1$ 
due to a naive numerical analysis~\cite{Apreda:2001us,Kobakhidze:2016mch}.
In this Letter,
we just remove such cases from our plot
because our conclusion is not affected.
The lower end of each black curve is drawn by this procedure.
The shaded regions represent the expected coverage
at the future space-based interferometers, 
eLISA~\cite{Klein:2015hvg,Caprini:2015zlo,PetiteauDataSheet} 
and DECIGO~\cite{Kawamura:2011zz}. 
The sensitivity regions of four eLISA detector configurations
described in Table I in Ref.~\cite{Caprini:2015zlo} are 
denoted by ``C1'', ``C2'', ``C3'' and ``C4''.
The expected sensitivities for 
the future DECIGO stages are labeled by
``Correlation'', ``1 cluster'' and ``Pre'' following 
Ref.~\cite{Kawamura:2011zz}.
Although the transition temperature $T_t$ depends on the HSM parameters,
we take $T_t^{}=50~\GeV$ for the purpose of illustration.
The experimental sensitivities are also dependent on the velocity
of the bubble wall $v_b^{}$, which is uncertain.
As a reference, we take $v_b=0.95$
so that strong GW signals are expected
\footnote{In Ref.~\cite{No:2011fi}, EWBG is not necessarily impossible
even in this case.}. 
If we take smaller values of $v_b^{}$ such as $0.2$, which the EWBG 
scenario prefers, the sensitivity area is pushed down to lower $\tilde{\beta}$
and larger $\alpha$ regions.
For $v_b^{}<1$,
the contribution from the sound waves is the dominant source of
the total GW spectrum while
those from the bubble wall collision and the turbulence are not significant~\cite{sw}.
This plot demonstrates that 
eLISA or DECIGO is capable of detecting
stochastic GWs from the sound wave source
in the most of the HSM parameter region with 1stOPT.

\begin{figure}[t]
  \centerline{\includegraphics[width=0.8\textwidth]{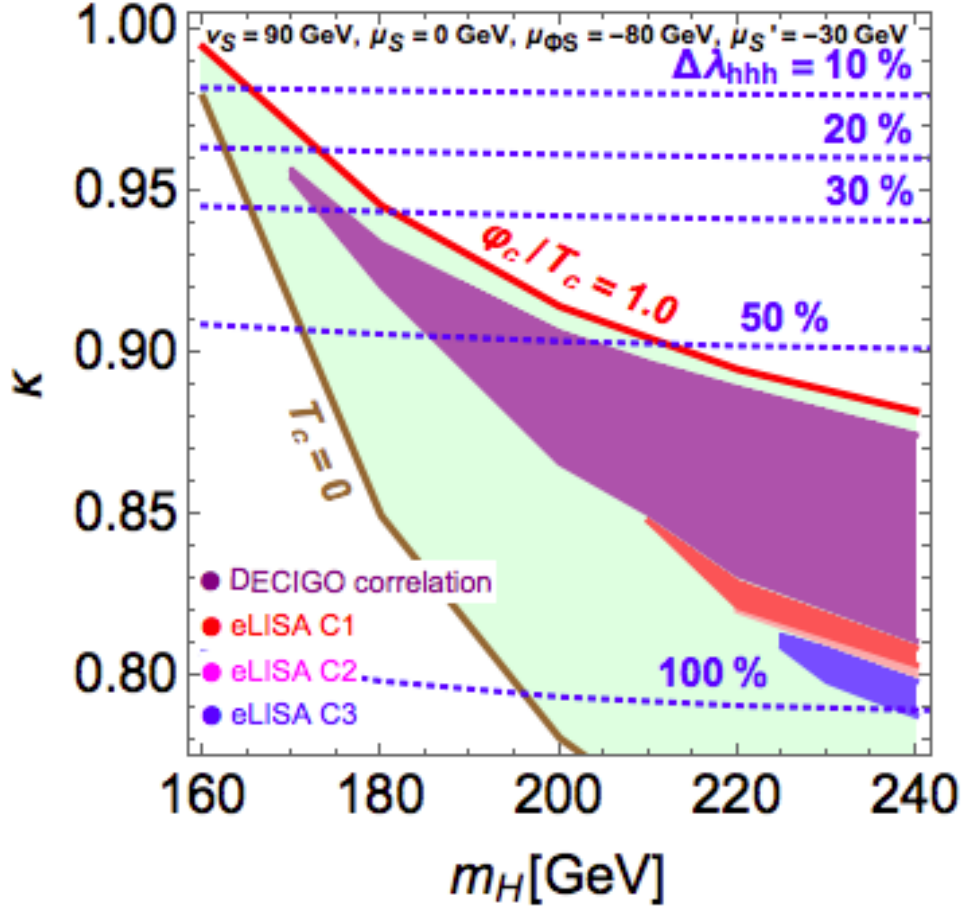}}
  \caption{The detectability of GWs and the contours of the deviations 
    in the $hhh$ coupling $\Delta \lambda_{hhh}^{}$ in the 
    $m_H^{}$-$\kappa$ plane.
    The projected region of a higher sensitive detector design 
    is overlaid with that of weaker one.
    The region which satisfies both $\varphi_c^{}/T_c> 1$ and $T_c^{}>0$
    is also shown for a reference.
    The input parameters and legends are same as in Fig.~\ref{fig:GW1}}
  \label{fig:GW2}
\end{figure}

In Fig.~\ref{fig:GW2}, the detectability of GWs 
and the contours of the deviation in the triple Higgs
boson coupling $\Delta \lambda_{hhh}^{}$ in the HSM are shown in
the $m_H$-$\kappa$ plane. 
The projected region of a higher sensitive detector design 
is overlaid with that of weaker one.
The region which satisfies both $\varphi_c^{}/T_c^{}> 1$ and $T_c^{}>0$
is also shown for a reference.
This plot highlights the importance of the synergy
between the precision measurements of the Higgs boson couplings
at future colliders
and the observation of stochastic GWs at future GW interferometers.
As deviations in the Higgs boson couplings from the SM values
are larger, the strength of 1stOPT and that of GW signals are more significant.
For example, 
once $\kappa$ is found to be smaller than 0.95 by LHC experiments, then the $hhh$ coupling should be greater than $20\%$.  
Such a deviation in the $hhh$ coupling can be measured at the ILC with $\sqrt{s}=1$~TeV~\cite{Fujii:2015jha}. 
In addition, we learn from Fig.~\ref{fig:GW2} that the scenario can also be well tested at DECIGO and eLISA. The combined measurements make it possible to identify the shape of the Higgs potential of the HSM. 

In this Letter, we have evaluated the spectrum of GWs that are generated
from the strongly 1stOPT of the EWSB in the HSM.
Based on the finite temperature one-loop effective potential
with the two scalar fields, 
the profile of the vacuum bubble and the transition temperature
for the tunneling from the false vacuum to the true one have been analyzed.
In view of EWBG, 
the parameter space allowed by the condition
of the departure from thermal equilibrium has been explored.
We have investigated the GW signals from
the sound waves, the bubble wall collision
and the turbulence resulting from the bubble collisions.
We have pointed out that the predicted peak GW amplitude 
from the sound wave contribution is so strong
as to be detected at
future space-based interferometers such as eLISA, DECIGO and BBO.
Deviations in the Higgs boson couplings have been also evaluated,
and found to be measurable at future colliders.
Therefore, we conclude that
the strongly 1stOPT of EWSB in the HSM
can be verified by combining
the precision measurements of various Higgs boson couplings at the LHC
and the $hhh$ coupling at future 
electron-positron colliders with those of
stochastic GW background at future space-based interferometers. 
\\

{\it Note Added:} During the completion of the manuscript, 
we became aware of an analogous calculation done independently 
by another group~\cite{Huang:2016cjm}.

\begin{acknowledgments}
This work was supported, in part, by 
Grant-in-Aid for Scientific Research on Innovative Areas,
the Ministry of Education, Culture, Sports, Science and Technology,
No.\ 16H01093 (MK) and No.\ 16H06492 (SK), 
Grant H2020-MSCA-RISE-2014~no.~645722 (Non Minimal Higgs) (SK),
and National Research Foundation
of Korea (NRF) Research Grant NRF-2015R1A2A1A05001869 (PK,TM).
\end{acknowledgments}


\end{document}